\documentclass[11pt,a4paper]{article}

\usepackage{bm,bbm,amsfonts,amssymb,mathrsfs,color}
\usepackage{graphicx}
\usepackage{hyperref}
\usepackage{amsmath,ascmac}

\newcommand{\Z}{\mathbb{Z}}
\newcommand{\C}{\mathbb{C}}

\newcommand{\N}{\mathcal{N}}

\newcommand{\calZ}{\mathcal{Z}}

\newcommand{\U}{\mathrm{U}}

\newcommand{\Sp}{\mathrm{Sp}}

\newcommand{\Tr}{\operatorname{Tr}}
\newcommand{\tr}{\operatorname{tr}}
\newcommand{\id}{\mathbbm{1}}

\makeatletter

\@addtoreset{equation}{section}
\makeatother

\date{\today}

\begin{document}

\begin{titlepage}

\renewcommand{\thefootnote}{\fnsymbol{footnote}}

\begin{flushright}
 {\tt 
 IPHT-T13/285 \\
 RIKEN-MP-82
 }
\\
\end{flushright}

\vskip7em

\begin{center}
 {\Large {\bf 
 Note on a duality of topological branes
 }}

 \vskip5em

 \setcounter{footnote}{1}
 {\sc Taro Kimura}\footnote{E-mail address: 
 \href{mailto:taro.kimura@cea.fr}
 {\tt taro.kimura@cea.fr}} 

 \vskip2em

{\it 
Institut de Physique Th\'eorique,
 CEA Saclay, F-91191 Gif-sur-Yvette, France
% \\ and \\
 \\ \vspace{.5em}
 Mathematical Physics Laboratory, RIKEN Nishina Center, Saitama 351-0198,
 Japan 
% \\ \vskip.2em 
% $^2$
}

 \vskip3em

% \today
\end{center}

 \vskip2em

\begin{abstract}
%We study the brane intersection in the topological B-model, which is
% realized by inserting both two kinds of the non-compact branes simultaneously.
We show a duality of branes in the topological B-model by inserting
 two kinds of the non-compact branes simultaneously.
We explicitly derive the integral formula for the matrix model partition
 function describing this situation, which correspondingly includes both
 of the characteristic polynomial and the external source.
We show that these two descriptions are dual to each other through the Fourier
 transformation, and the brane partition function satisfies integrable
 equations in one and two dimensions.
\end{abstract}

\end{titlepage}

\tableofcontents

\setcounter{footnote}{0}

%%%%%%%%%%   Introduction   %%%%%%%%%%

\section{Introduction}\label{sec:intro}

D-brane is the most fundamental, but also non-perturbative object in
superstring theory. 
It gives a non-trivial boundary condition for open strings, and
consequently a lot of gauge theories can be constructed as effective
theories for stringy modes on them.
% by considering combinations of branes.
When we would implement various gauge groups and matter contents for the
effective field theory, branes often have to be arranged in a
complicated way.
In such a case, we have to deal with dynamics of brane intersection
appropriately, and discuss what kind of stringy excitation can be
allowed there.
In this sense it is important to study properties of a brane complex
in detail.

In this article we investigate some aspects of brane-on-brane structure in
topological string theory by introducing two kinds of branes
simultaneously, especially through its matrix model description.
It is well known that the Riemann surface plays a crucial role in
the topological B-model on Calabi--Yau threefolds, and this
Riemann surface can be identified with the spectral curve appearing in the
large $N$ limit of the matrix model~\cite{Dijkgraaf:2002fc}.
A brane is also introduced into the B-model,
and its realization in terms of the matrix model was
extensively investigated~\cite{Aganagic:2003qj}:
There are seemingly two kinds of non-compact branes, which correspond to
the characteristic polynomial and the external source in the matrix model.
While these two descriptions are apparently different, we will see
that they are essentially the same, as a consequence of the symplectic
invariance of the topological B-model.
%By considering both kinds of the non-compact branes at once, we can
%discuss intersection of branes in the topological B-model.
From the viewpoint of the matrix model, this situation just corresponds to
insertion of both of the characteristic polynomial and the external
source.
We will explicitly show that two descriptions are dual to each other by
seeing a newly derived matrix integral formula.

It is known that the correlation function of the characteristic
polynomial and the matrix integral with the external source are
deeply related to the integrable hierarchy~\cite{Morozov:1994hh}.
Thus it is naively expected that the hybrid of them also possesses
a similar connection to the integrability.
Since in this case we have two kinds of variables corresponding to the
characteristic polynomials and the external sources, we obtain the
two dimensional integrable equation as a consequence, and identify the
partition function as the $\tau$-function of the corresponding
integrable hierarchy with the Miwa coordinates.

This article is organized as follows.
In Sec.~\ref{sec:brane} we start with a review on how to consider branes
in the topological B-model in terms of the matrix model.
We introduce two descriptions of branes using the characteristic
polynomial and the external source of the matrix integral with emphasis
on their similarity and difference between them.
In Sec.~\ref{sec:intersection} we show a duality between two kinds of
non-compact branes in the B-model by introducing both of them at the
same time.
We derive a new matrix integral formula corresponding to this situation,
and show that the characteristic polynomial and the external source
are dual to each other in the sense of Fourier transformation.
We also show that the brane partition function satisfies
the integrable equation in one and two dimensions, which is well-known
as the Toda lattice equation.
In Sec.~\ref{sec:Gaussian} we consider the Gaussian matrix
model as an example.
In this case we can show the duality using the
fermionic variables, which is useful to discuss the effective degrees of
freedom on the branes.
We close this article in Sec.~\ref{sec:discussion} with some discussions
and concluding remarks.

\section{Branes in the topological B-model}\label{sec:brane}

We first review the large $N$ matrix model description of the
topological B-model, mainly following~\cite{Aganagic:2003qj}, and how to
introduce the brane, which is also called the B-brane. %or the FZZT-brane.
The non-compact Calabi--Yau threefold we discuss in this article is
obtained as a hypersurface in $(u, v, p, x)\in\C^4$,
\begin{equation}
% \mathrm{CY} : \
 u \, v - H (p, x) = 0 
  \, .
  \label{CY01}
\end{equation}
Here $H(p,x)$ determines the Riemann surface
\begin{equation}
 \Sigma : \ H(p,x) = 0 
  \label{RS01}
\end{equation}
with
\begin{equation}
  H(p,x) = p^2 - W'(x)^2 - f(x) \, ,
   \label{Ham01}
\end{equation}
where $W(x)$ and $f(x)$ are polynomials of degree
$n+1$ and $n$, respectively.
This smooth Calabi--Yau threefold is given by deformation of the singular one
\begin{equation}
 H(p, x) = p^2 - W'(x)^2
  \, . 
\end{equation}
The branes, which are compact, are wrapping $n$ $S^3$ %$\mathbf{P}^1$ 
at critical points $W'(x) = 0$, and the sizes of $S^3$'s s%$\mathbf{P}^1$'s 
are parametrized by $f(x)$, which describes the quantum correction
around them.

The holomorphic $(3,0)$-form $\Omega$ in the Calabi--Yau threefold
(\ref{CY01}) is chosen to be
\begin{equation}
 \Omega = \frac{du \wedge dp \wedge dx}{u}
  \, .
  \label{(3,0)-form}
\end{equation}
Then the periods of this $(3,0)$-form $\Omega$ over three-cycles reduce
to the integral of the symplectic 2-form
\begin{equation}
 \int_D dp \wedge dx
  \, ,
  \label{symplectic-form}
\end{equation}
where $D$ stands for domains in the complex two-dimensional
$(p,x)$-plane.
The boundary of this domain denoted by $\gamma$ is related to the
Riemann surface (\ref{RS01}) as $\gamma = \partial D \subset \Sigma$.
Thus this integral further reduces to a one-cycle on the Riemann surface
$\Sigma$
\begin{equation}
 \int_\gamma p \, dx
  \, .
\end{equation}
In this way we can focus only on the complex one-dimensional subspace $\Sigma$
%the topological B-model given as a quantization of the variation of
%complex structures on the Calabi--Yau threefold
by keeping the dependence on $u$ and $v$ fixed.

\subsection{Characteristic polynomial}

The algebraic curve (\ref{RS01}) can be identified
with the spectral curve of the matrix model:
It is just given by the loop equation in the large $N$ limit of the
matrix integral
\begin{equation}
 \calZ_N = \int \! dX \, e^{-\frac{1}{g_s}\Tr W(X)} 
   \, .
\end{equation}
The polynomial $W(x)$ is the matrix potential.
This is the reason why we can discuss the topological B-model using the
matrix model. 
In this description the genus expansion with $1/N$ corresponds to the
quantum correction, and the 't Hooft parameter $t = g_s N$ gives the
size of $S^3. $%$\mathbf{P}^1$.

The other canonical variable in (\ref{Ham01}) is related to the
resolvent of the matrix model
\begin{equation}
 p(x) = W'(x) - 2 g_s \Tr \frac{1}{x-X} \, .
\end{equation}
The saddle point equation of the matrix model is equivalent to the
condition $p(x) = 0$.
Since the one-form on the spectral curve is given by these two canonical
variables
\begin{equation}
 \lambda = p(x) \, dx = d \phi \, ,
  \label{1form}
\end{equation}
we can naturally introduce the chiral boson $\phi(x)$ on the Riemann surface
$\Sigma$, which is interpreted as the Kodaira--Spencer field describing
deformation of the complex structure at infinity
\begin{equation}
 \phi(x) = W(x) - 2 g_s \Tr \log (x - X) \, .
  \label{eff_pot}
\end{equation}
Here this $\phi(x)$ also has the meaning of the effective potential for the
matrix model, and thus the eigenvalues are degenerated at the critical
point, $\partial \phi = p(x) = 0$.

The vertex operator, which creates a non-compact brane at a position
$x$, is constructed by the standard bosonization scheme
\begin{equation}
 V(x) =  e^{-\frac{1}{2g_s}\phi(x)} = e^{-\frac{1}{2g_s}W(x)} \det(x-X) \, .
  \label{vertex_op}
\end{equation}
The prefactor $e^{-\frac{1}{2g_s}W(x)}$ corresponds to the classical
part of the operator, while the determinant part, namely the
characteristic polynomial,
%$\det (x - X)$, 
gives the quantum fluctuation as a gravitational back reaction.
%This is the first way of describing the non-compact brane in the B-model.
This brane creation operator gives a pole at $x$ on the Riemann surface, and
its residue is given by
\begin{equation}
 \oint \lambda = g_s \, .
  \label{BohrSommerfeld}
\end{equation}
This is just interpreted as one brane contribution, and also shows that
$g_s$ plays a role of the Planck constant $\hbar$ for the canonical pair
$(p,x)$.

The partition function of the branes is represented as a correlation
function of the characteristic polynomials,
e.g., $k$-point function given by
\begin{equation}
 \left\langle
  \prod_{\alpha=1}^k \det (x_\alpha - X)
 \right\rangle \, .
\end{equation}
This expectation value is taken with respect to the matrix measure
\begin{equation}
 \Big\langle
  \mathcal{O}(X)
 \Big\rangle
 = 
 \frac{1}{\calZ_N}
 \int \! dX \, \mathcal{O}(X) \, e^{-\frac{1}{g_s}\Tr W(X)} 
 \, 
\end{equation}
with the standard normalization $\langle \, 1 \, \rangle = 1$.
Including the classical part, the brane partition function is then given by
\begin{equation}
 \Psi_k (x_1, \cdots, x_k)
  = 
  \prod_{\alpha=1}^k e^{-\frac{1}{2g_s} W(x_\alpha)}
 \left\langle
  \prod_{\alpha=1}^k \det (x_\alpha - X)
 \right\rangle 
 \, .
\end{equation}
This correlation function can be exactly evaluated using the orthogonal
polynomial method.
We will come back to this formula in Sec.~\ref{sec:integrability} (see
(\ref{ch_poly01})).

\subsection{External source}\label{sec:ext_source}

Let us then consider another kind of the non-compact brane in the
B-model, which is described by the external source in the matrix model.
%One of the most important example of the matrix model with the external
%source is the Kontsevich model \cite{Kontsevich:1992ti}.
We consider the matrix action written in a form of 
 \begin{equation}
 S(P) = \frac{1}{g_s} \Tr
  \Big[
   W(P) - A P
  \Big] \, ,
  \label{mat_action01}
\end{equation}
where the potential is regarded as an integral of the one-form along an
open path to a certain point $p$ on the Riemann surface
\begin{equation}
 W(p) = \int^p \! x(p') \, dp' \, .
  \label{mat_pot_int}
\end{equation}
We can assume the matrix $A$ is diagonal $A = \mathrm{diag} ( a_1,
\cdots, a_N)$ without loss of generality.
The action (\ref{mat_action01}) corresponds to the matrix model with the
external source
\begin{equation}
 \calZ_N (A)
  =
  \int \! dP \,
  e^{-\frac{1}{g_s} \Tr \Big[ W(P) - A P \Big]}
  \, ,
\end{equation}
which is analogous to the Kontsevich model~\cite{Kontsevich:1992ti}.
The external source implies positions of $N$ branes, at least
at the classical level, because the extremum of the action
$W'(P) - A = 0$ gives the classical solution, $X =
\mathrm{diag}(a_1, \cdots, a_N)$, since we have $W'(P) = X$ according to
(\ref{mat_pot_int}).

The one-form used in (\ref{mat_pot_int}) is apparently different from
(\ref{1form}), but they are equivalent up to the symplectic
invariance for a pair of the canonical variables $(p,x)$.
This symmetry is manifest by construction of the topological B-model as
seen in (\ref{(3,0)-form}) and (\ref{symplectic-form}).
Therefore two descriptions of the non-compact branes based on the
characteristic polynomial and the external source in the matrix model
are dual to each other in this sense.
We will show in Sec.~\ref{sec:intersection} that they are converted through
the Fourier transformation by deriving the explicit matrix integral
representation.

\section{A duality of branes}\label{sec:intersection}

As seen in the previous section, there are two kinds of non-compact
branes in the topological B-model, which are related through the
symplectic transformation.
We then study the situation such that both kinds of branes are
applied at once.
This is realized by inserting the characteristic polynomial to the matrix
model in addition to the external source.
Let us consider the corresponding partition function denoted by
\begin{equation}
% \left\langle
%  e^{\Tr AX}
%  \prod_{\alpha=1}^M
%  \det (\lambda_\alpha - X)
% \right\rangle
 \Psi_{N,\, M} 
 \left( \{a_j\}_{j=1}^N; \{\lambda_\alpha\}_{\alpha=1}^M \right)
 =
 \int \! dX \, e^{-\frac{1}{g_s} \Tr W(X) + \Tr AX}
  \prod_{\alpha=1}^M
  \det (\lambda_\alpha - X)
  \, .
  \label{dual_part_func}
\end{equation}
We now have to be careful of the meaning of the matrix
potential and the external source.
As discussed in Sec.~\ref{sec:ext_source}, if the matrix potential is
given by the integral of the one-form in the form of
(\ref{mat_pot_int}), the external source gives classical positions
of branes in the $x$-coordinate.
In this case, however, the roles of $x$ and $p$ are exchanged.
Instead of (\ref{mat_pot_int}), the potential should be written as
\begin{equation}
 W(x) = \int^x \! p(x) \, dx \, .
\end{equation}
With this choice, the corresponding external source determines positions
of branes in the $p$-coordinate.
We note that, in this case, the extremum of the action does not simply
imply the classical positions of branes as $W'(X) = A$, because of
potential shift due to the characteristic polynomial.
Although there exists this kind of back reaction, the interpretation of
the external source as the $p$-coordinate still holds as shown in the
following.

\begin{figure}[t]
 \begin{center}
  \includegraphics[width=40em]{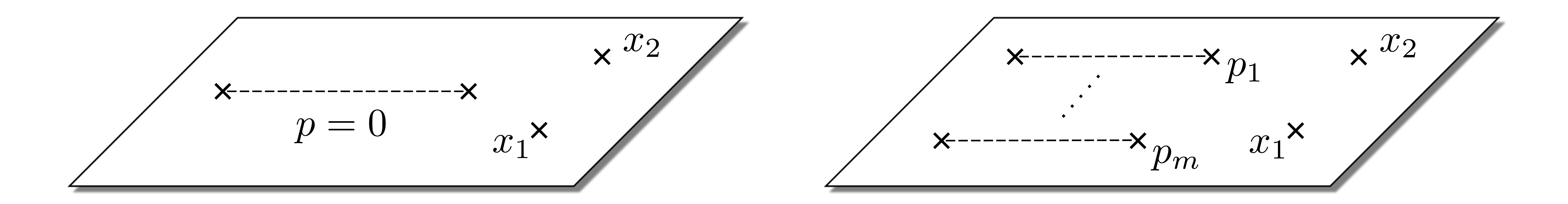}
 \end{center}
 \caption{%Insertion of branes from two directions, $x$ and $p$.
% which are intersecting on the Riemann surface $\Sigma$.
% Branes extended along each direction are created by the characteristic
% polynomials and the external sources, respectively.
 (Left) In the case with $A=0$, the one-cut matrix model has a single cut in
 the complex $x$-plane, which is seen as a pole at $p=0$ with degree
 $N$ in the $p$-plane.
 There are also simple poles created by the characteristic polynomials.
 (Right) When one applies the external source $A = \mathrm{diag}(p_1, \cdots,
 p_1, \cdots, p_m, \cdots, p_m)$, it is split into $m$ distinct
 poles with the corresponding multiplicity.
 }
 \label{sphere}
\end{figure}

The characteristic polynomial and the external source put punctures on the
Riemann surface $\Sigma$ for $x$ and $p$ coordinates because they
corresnpond to the vertex operators creating the non-compact branes.
Since $(p,x)$ is a pair of the canonical variables, if once its
$x$-coordinate is fixed, one cannot determine the other $p$-coodinate, and
vice versa.
This means that the positions of the branes created by the
characteristic polynomial and the external source are fixed by the $x$
and $p$ coordinates.
For example, in the case with $A = 0$ for simplicity, all the
$N$ eigenvalues are distributed on the extended line, namely the cut in
the complex $x$-plane described by $p(x) = 0$, which corresponds to the
one-form $p(x)dx$.
On the other hand, if we apply the other representation based on the
$p$-plane, the one-form $x(p)dp$ has a pole at $p = 0$ with degree $N$.
Then, when one turns on the external source $A = \mathrm{diag}(a_1, \cdots,
a_N)$, the pole at $p = 0$ is split and located at $p = a_i$ for
$i = 1, \cdots, N$.
In this way the external source, and also the characteristic polynomial,
characterize a pole with the $p$ and $x$ coordinates, respectively.
% and thus they extended along $p$ and $x$ directions, respectively.
%Since their positions are labeled by the $x$ and $p$ coordinates,
%respectively, these branes are extended along perpendicular directions.
In Fig.~\ref{sphere} we illustrate the situation such that both kinds of
branes are inserted to the Riemann surface $\Sigma$.
%We call this the brane intersection in the topological B-model.
We will see in the following that these coordinates can be exchanged
through the Fourier transformation.

\subsection{Matrix integral formula}\label{sec:formula}

We then provide an explicit formula for the partition function
(\ref{dual_part_func}).
In order to derive the formula, we apply a method to compute
the matrix model partition function with the external
source~\cite{Dijkgraaf:1991qh,Kharchev:1991cy}, which is also
applicable to this situation.

First of all, we move to ``eigenvalue representation'' from the
$N \times N$ Hermitian matrix integral (\ref{dual_part_func}) by
integrating out the angular part of $X$. 
This can be done by using Harish-Chandra--Itzykson--Zuber
formula~\cite{HC:1957,Itzykson:1979fi}
\begin{equation}
 \int \! dU \, e^{\Tr U X U^\dag Y}
  =
  \frac{\det e^{x_i y_j}}{\Delta(x) \Delta(y)}
  \, ,
\end{equation}
where the integral is taken over $\U(N)$ group with the Haar measure
normalized by the volume factor $\operatorname{Vol}(\U(N))$, and $\Delta(x)$ is 
the Vandermonde determinant.
Thus we have the eigenvalue representation of (\ref{dual_part_func})
\begin{equation}
 \int \! d^N x \,
  \frac{\Delta(x)}{\Delta(a)}
  \prod_{j=1}^N e^{-\frac{1}{g_s} W(x_j) + a_j x_j}
  \prod_{j=1}^N \prod_{\alpha=1}^M 
  (\lambda_\alpha - x_j)
  \, .
\end{equation}

We here apply the formula
\begin{equation}
 \Delta(x) \Delta(\lambda)
  \prod_{\alpha=1}^M \prod_{j=1}^N 
%  \prod_{j,\alpha} 
%  \prod_{\substack{1 \le j \le N \\ 1 \le \alpha \le M}}
  (\lambda_\alpha - x_j)
  = \Delta(x,\lambda)
  \, .
\end{equation}
The RHS of this equation $\Delta(x,\lambda)$ is the Vandermonde
determinant for $N+M$ variables $(x_1, \cdots, x_N, \lambda_1, \cdots,
\lambda_M)$, which is represented as an $(N+M) \times (N+M)$ determinant
\begin{equation}
 \Delta(x,\lambda)
  =
  \det
  \left(
   \begin{array}{cc}
    x_k^{j-1} & x_k^{N+\alpha-1} \\
    \lambda_\beta^{j-1} & \lambda_\beta^{N+\alpha-1} \\
   \end{array}
  \right)
  =
  (-1)^{MN}
  \det
  \left(
   \begin{array}{cc}
    \lambda_\beta^{\alpha-1} & \lambda_\beta^{M+j-1} \\
    x_k^{\alpha-1} & x_k^{M+j-1} \\
   \end{array}
  \right)
  \, .
  \label{det_N+M}
\end{equation}
We recall that $j, k = 1, \cdots, N$ and $\alpha, \beta = 1,
\cdots, M$.
Since each matrix element can be replaced with any monic polynomials, $P_k(x)
= x^k + \cdots$, this determinant (\ref{det_N+M}) is written in a more
general form
\begin{equation}
 \Delta(x,\lambda)
  =
  \det
  \left(
   \begin{array}{cc}
    P_{j-1}(x_k) & P_{N+\alpha-1}(x_k) \\
    P_{j-1}(\lambda_\beta) & P_{N+\alpha-1}(\lambda_\beta) \\
   \end{array}
  \right)
  =
  (-1)^{MN}
  \det
  \left(
   \begin{array}{cc}
    P_{\alpha-1}(\lambda_\beta) & P_{M+j-1}(\lambda_\beta) \\
    P_{\alpha-1}(x_k) & P_{M+j-1}(x_k) \\
   \end{array}
  \right)
  \, .
  \label{det_N+M02}
\end{equation}
Thus the integral is given by
\begin{equation}
 \frac{1}{\Delta(a) \Delta(\lambda)}
  \int \! d^N x \,
  \Delta(x,\lambda)
  \prod_{j=1}^N e^{-\frac{1}{g_s} W(x_j) + a_j x_j}
  \, .
\end{equation}
To perform this integral, we then introduce an auxiliary function
\begin{equation}
 Q_k(a)
  = 
  \int \! dx \, 
  P_k(x) \,
  e^{-\frac{1}{g_s} W(x) + a x}
  \, .  
  \label{Q_func}
\end{equation}
Using this function, we arrive at the final expression of the
partition function (\ref{dual_part_func})
\begin{eqnarray}
  \Psi_{N,\, M} \left( \{a_j\}_{j=1}^N; \{\lambda_\alpha\}_{\alpha=1}^M \right)
  & = &
  \frac{1}{\Delta(a) \Delta(\lambda)}
  \det
  \left(
   \begin{array}{cc}
    Q_{j-1}(a_k) & Q_{N+\alpha-1}(a_k) \\
    P_{j-1}(\lambda_\beta) & P_{N+\alpha-1}(\lambda_\beta) \\
   \end{array}
  \right)
  \nonumber \\
 & = &
   \frac{(-1)^{MN}}{\Delta(a) \Delta(\lambda)}
  \det
  \left(
   \begin{array}{cc}
    P_{\alpha-1}(\lambda_\beta) & P_{M+j-1}(\lambda_\beta) \\
    Q_{\alpha-1}(a_k) & Q_{M+j-1}(a_k) \\
   \end{array}
  \right)
  \, .
  \label{dual_part_func02}
\end{eqnarray}
This expression is manifestly symmetric under
the exchange of $(a_1, \cdots, a_N)$ and $(\lambda_1, \cdots, \lambda_M)$
with the transformation: $P_k(\lambda) \leftrightarrow Q_k(a)$.
As seen in (\ref{Q_func}), this is nothing but a Fourier (Laplace)
transform of $x^k e^{-\frac{1}{g_s}W(x)}$.
Thus we can say that the characteristic polynomial and the external source in
the matrix model are dual to each other in the sense of Fourier
transformation.

In terms of the topological strings, this duality reflects the
symplectic invariance of the canonical pair $(p,x)$ in the B-model, as
seen in (\ref{symplectic-form}).
Thus such a symmetry, which allows to exchange the canonical variables $x$
and $p$, is directly related to the duality for the $(r,s)$
minimal model described by $H(p,x) = p^r + x^s + \cdots$, and
also the open/closed string
duality~\cite{Gaiotto:2003yb,Maldacena:2004sn,Hashimoto:2005bf}.
It should be also mentioned that the symplectic invariance, or
$\mathrm{SL}(2,\Z)$ symmetry exchanging the canonical pair, realizes
the S-duality of the topological string or topological
M-theory~\cite{Dijkgraaf:2004te}.
Since the two descriptions of the non-compact branes are just converted
into each other through the Fourier transformation, they are essentially
equivalent.
Although this equivalence has been already pointed out for the Gaussian
matrix model, as in~\cite{Aganagic:2003qj}, we can see that this
symmetry even holds with a generic matrix potential $W(x)$.
This matrix potential determines the geometry of the Calabi--Yau
threefold in the form of (\ref{Ham01}).
In addition, the positions of the branes are in a relationship
satisfying the uncertainty principle:
one cannot determine both of the $x$ and $p$ coordinates at the same
time, but only either of them.
%This also supports the interpretation of (\ref{dual_part_func}) as the
%intersecting brane partition function.
We also note that this kind of symplectic invariance appears quite
generally in the topological expansion of the spectral
curve~\cite{Eynard:2007kz,Eynard:2007nq,Eynard:2013csa}.

\subsection{Integrability}\label{sec:integrability}

The formula (\ref{dual_part_func02}) is a quite natural generalization
of the well-known formulae for
the expectation value of characteristic polynomial product
\begin{equation}
 \left\langle
  \prod_{\alpha=1}^M \det (\lambda_\alpha - X)
 \right\rangle
 =
 \frac{1}{\Delta(\lambda)} 
 \det_{1 \le \alpha, \beta \le M} P_{N+\alpha-1}(\lambda_\beta)
 \, ,
 \label{ch_poly01}
\end{equation}
where $P_k(x)$ is $k$-th monic orthogonal polynomial with respect to the
weight function $w(x) = e^{-\frac{1}{g_s}W(x)}$,
and also the matrix integral with the external source
\begin{equation}
 \int \! dX \, e^{-\frac{1}{g_s} \Tr W(X) + \Tr AX}
  =
  \frac{1}{\Delta(a)}
  \det_{1 \le j, k \le N} Q_{j-1}(a_k)
  \, .
  \label{ext_source01}
\end{equation}
It is convenient to apply the simplest choice of the polynomial
$P_k(x)=x^k$ to this formula (\ref{ext_source01}).
In this case the function $Q_k(a)$ is given by
\begin{equation}
 Q_k (a)
  = 
  \int \! dx \, 
  x^k \,
  e^{-\frac{1}{g_s} W(x) + a x}
  =
  \left( \frac{d}{da} \right)^k Q(a)
  \, ,
  \label{aux_func02}
\end{equation}
with an Airy-like function
\begin{equation}
 Q (a)
  = 
  \int \! dx \, 
  e^{-\frac{1}{g_s} W(x) + a x}
  \, .
  \label{aux_func01}
\end{equation}
See \cite{Morozov:1994hh} and references therein for details

It is known that this kind of determinantal formula generically
plays a role as the $\tau$-function~\cite{Kharchev:1991cy}, and
satisfies the Toda lattice equation by taking the equal parameter
limit.
See, for example,~\cite{Verbaarschot:2005rj}.
We show that the formula (\ref{dual_part_func02}) indeed satisfies a similar
integrable equation in the following.

Let us parametrize the positions of branes by ``center of mass'' and
deviations from it as $a_j = a + \delta a_j$ and 
$\lambda_\alpha = \lambda + \delta \lambda_\alpha$.
We rewrite the numerator in terms of the deviations $\{\delta
a_j\}$ and $\{\delta \lambda_\alpha\}$ by considering the Taylor expansion
around the center of masses
\begin{equation}
  \det
  \left(
   \begin{array}{cc}
    Q_{j-1}(a_k) & Q_{N+\alpha-1}(a_k) \\
    P_{j-1}(\lambda_\beta) & P_{N+\alpha-1}(\lambda_\beta) \\
   \end{array}
  \right)
  =
  \det
  \left(
   \begin{array}{cc}
    \frac{(\delta a_k)^{l-1}}{(l-1)!} & \\
    & \frac{(\delta \lambda_\beta)^{\gamma-1}}{(\gamma-1)!} \\
   \end{array}
  \right)
  \det
  \left(
   \begin{array}{cc}
    Q_{j-1}^{(l-1)}(a) & Q_{N+\alpha-1}^{(l-1)}(a) \\
    P_{j-1}^{(\gamma-1)}(\lambda) & P_{N+\alpha-1}^{(\gamma-1)}(\lambda) \\
   \end{array}
  \right)
  \, ,
\end{equation}
where 
$\displaystyle P_{j-1}^{(\gamma-1)}(\lambda) = \left(\frac{d}{d
\lambda}\right)^{\gamma-1} P_{j-1}(\lambda)$, 
$\displaystyle Q^{(l-1)}_{j-1}(a) = \left(\frac{d}{da}\right)^{l-1}
Q_{j-1}(a) = Q_{j+l-2}(a)$ and so on.
The first determinant in the RHS is almost canceled by the Vandermonde
determinants in the denominator of (\ref{dual_part_func02}), since they
are invariant under the constant shift as $\Delta(a) = \Delta(\delta a)
= \det (\delta a_k)^{j-1}$ and $\Delta(\lambda) = \Delta(\delta \lambda)
= \det (\delta \lambda_\beta)^{\alpha-1}$, respectively.
Therefore the partition function in the equal position limit becomes
\begin{eqnarray}
 \Psi_{N,\, M} 
  \left(
   \{ a_j = a \}_{j=1}^N ;
   \{ \lambda_\alpha = \lambda \}_{\alpha=1}^M
  \right)
  & = &
  c_{N,\,M}
  \det
  \left(
   \begin{array}{cc}
    Q_{j-1}^{(k-1)}(a) & Q_{N+\alpha-1}^{(k-1)}(a) \\
    P_{j-1}^{(\beta-1)}(\lambda) & P_{N+\alpha-1}^{(\beta-1)}(\lambda) \\
   \end{array}
  \right)
  \label{equal_pos01} \\
 & = &
  (-1)^{MN}
  c_{N,\,M}
  \det
  \left(
   \begin{array}{cc}
    P_{\alpha-1}^{(\beta-1)}(\lambda) & P_{M+j-1}^{(\beta-1)}(\lambda) \\
    Q_{\alpha-1}^{(k-1)}(a) & Q_{M+j-1}^{(k-1)}(a) \\
   \end{array}
  \right)
  \nonumber \\
  \label{equal_pos02} \\
 & \equiv &
  c_{N,\,M}
  \det R_{N,\, M}(a,\lambda)
  \, ,
  \nonumber
\end{eqnarray}
with
\begin{equation}
 c_{N,\, M}
  =
  \prod_{j=1}^{N} \frac{1}{(j-1)!}
  \prod_{\alpha=1}^{M} \frac{1}{(\alpha-1)!} \, ,
\end{equation}
which corresponds to the volume factor for $\U(N)$ and $\U(M)$ group.
This expression is seen as a hybridized version of
the Wronskian.
In the following we apply $P_k(x) = x^k$ and (\ref{aux_func02}) as
the case of (\ref{ext_source01}) for simplicity.

In order to derive the integrable equation, we now use the Jacobi
identity for determinants, which is given by
\begin{equation}
 D \cdot D \left(
	    \begin{array}{cc}
	     i & j \\
	     k & l \\
	    \end{array}
	   \right)
 =
 D \left(
    \begin{array}{c}
     i \\ k
    \end{array}
   \right)
 \cdot
 D \left(
    \begin{array}{c}
     j \\ l
    \end{array}
   \right)
 -
 D \left(
    \begin{array}{c}
     i \\ l
    \end{array}
   \right)
 \cdot
 D \left(
    \begin{array}{c}
     j \\ k
    \end{array}
   \right)
 \, ,
 \label{Jacobi_id}
\end{equation}
where $D$ is a determinant, and the minor determinant
$D \left(\begin{array}{c} i \\ j \end{array} \right)$ is obtained by
removing $i$-th row and $j$-th column from the matrix.
Similarly 
$D \left(\begin{array}{cc} i & j \\ k & l \\ \end{array} \right)$ is obtained by
eliminating $i, j$-th row and $k,l$-th column.
Putting $i = k = N + M$ and $j = l = N + M - 1$ for the determinant in
(\ref{equal_pos02}), we have
\begin{eqnarray}
  \det R_{N,\,M} \cdot \det R_{N-2,\, M}
  & = &
  \det R_{N-1,\, M} \cdot M \lambda \partial_a^2 \det R_{N-1,\, M} 
  \nonumber \\
 && 
  - \, \partial_a \det R_{N-1,\, M} \cdot 
  M \lambda \partial_a \det R_{N-1,\,M}
  \, . 
  \label{R_relation01}
\end{eqnarray}
Here we have used a formula (\ref{shift_det}) discussed in
Appendix~\ref{sec:app}.
This provides the following relation for the equal position
partition function
\begin{equation}
 \frac{\Psi_{N+1,\,M} \cdot \Psi_{N-1,\,M}}{\left(\Psi_{N,\,M}\right)^2}
  =
  \frac{M}{N} \lambda \frac{\partial^2}{\partial a^2} \log \Psi_{N,\,M}
  \, .
  \label{Toda_eq01}
\end{equation}
This is just the Toda lattice equation along the $a$-direction, but with
a trivial factor which can be removed by rescaling the function.

We can assign another relation to the partition function by 
the identity (\ref{Jacobi_id}) for the expression (\ref{equal_pos01}) with
$i = k = N + M$ and $j = l = N + M - 1$, which reads
\begin{eqnarray}
  \det R_{N,\,M} \cdot \det R_{N,\, M-2}
  & = &
  \det R_{N,\, M-1} \cdot (M-1) 
  \partial_a \partial_\lambda \lambda \det R_{N,\, M-1} 
  \nonumber \\
 && 
  - \, \partial_\lambda \det R_{N,\, M-1} \cdot 
  (M-1) \lambda \partial_a \det R_{N,\,M-1}
  \, .
  \label{R_relation02}
\end{eqnarray}
We have again used the relation (\ref{shift_det}).
Rewriting this relation in terms of the partition function
(\ref{equal_pos01}), we have
\begin{equation}
 \frac{\Psi_{N,\,M+1} \cdot \Psi_{N,\,M-1}}{\left(\Psi_{N,\,M}\right)^2}
  =
  \frac{\partial^2}{\partial a \partial \lambda} \lambda \log \Psi_{N,\,M}
  \, .
  \label{Toda_eq02}
\end{equation}
We then obtain the two-dimensional Toda lattice
equation~\cite{Mikhailov:1979} with an extra factor.
In order to remove this irrelevant factor, we rescale the partition function
\begin{equation}
 \tilde{\Psi}_{N,\,M}(a,\lambda) 
  = e^{-\lambda} \, \Psi_{N,\,M}(a,\lambda)
  \, .
  \label{rescaled}
\end{equation}
Thus the Toda lattice equations (\ref{Toda_eq01}) and
(\ref{Toda_eq02}) are now written in the well-known form%
\footnote{
For example, see the reference~\cite{Hirota:1988} for the bilinear form
of the Toda lattice equation with the $\tau$-functions.
}
\begin{equation}
 \frac{\tilde\Psi_{N+1,\,M} \cdot \tilde\Psi_{N-1,\,M}}
      {\left(\tilde\Psi_{N,\,M}\right)^2}
  =
  \frac{M}{N} \frac{\partial^2}{\partial a^2} \log \tilde \Psi_{N,\,M}
  \, , \quad
 \frac{\tilde\Psi_{N,\,M+1} \cdot \tilde\Psi_{N,\,M-1}}
      {\left(\tilde\Psi_{N,\,M}\right)^2}
  =
  \frac{\partial^2}{\partial a \partial \lambda} \log \tilde\Psi_{N,\,M}
  \, .
\end{equation}
This means that the brane partition function
(\ref{rescaled}) plays a role of the $\tau$-function for the
one-dimensional, and also the two-dimensional Toda lattice equation
simultaneously.
We note that this is an exact result for finite $N$ (and also $M$).
If one takes the large $N$ limit, corresponding to the continuum limit for the
Toda lattice equations, it reduces to the KdV/KP equations.
We also comment that the $\tau$-function of the two-dimensional Toda
lattice hierarchy can be realized as the two-matrix
model integral~\cite{Adler:1997az,Adler:1997hq}.

Although, in this section we have focused only on the equal position
limit of the partition function (\ref{dual_part_func02}), it can be
regarded as the $\tau$-function for the corresponding integrable hierarchy.
In this case we can introduce two kinds of the Miwa coordinates
\begin{equation}
 t_n
  = \frac{1}{n} \, 
  \Tr A^{-n}
%  \sum_{j=1}^N \frac{1}{a_j^n}
  \, , \qquad
 \tilde{t}_n 
  = \frac{1}{n} \, 
  \tr \Lambda^{-n}
%  \sum_{\alpha=1}^M \frac{1}{\lambda_\alpha^n}
  \, .
\end{equation}
It is shown that all the time variables, $t_n$ and $\tilde{t}_n$, are
trivially related to each other in the equal parameter limit.
After taking the continuum limit, namely the large $N$ limit of the
matrix model, it shall behave as the $\tau$-function for the KdV/KP
hierarchies.

\section{Gaussian matrix model}\label{sec:Gaussian}

We now study a specific example of the matrix model with the harmonic
potential $W(x) = \frac{1}{2} x^2$, namely the Gaussian matrix model.
In this case we can check the duality
formula more explicitly~\cite{Brezin:2000CMP,Brezin:2007aa,Desrosiers:2008tp}:
\begin{equation}
 \frac{1}{\calZ_N}
  \int \! dX \,
  e^{-\frac{1}{2g_s} \Tr (X-A)^2}
  \prod_{\alpha=1}^M \det (\lambda_\alpha - X)
  =
  (-1)^{MN}
  \frac{1}{\calZ_M}
  \int \! dY \,
  e^{-\frac{1}{2g_s} \tr (Y-i\Lambda)^2}  
  \prod_{j=1}^N \det (a_j + i Y)
  \, ,
  \label{Gaussian_duality}
\end{equation}
where $X$ and $Y$ are Hermitian matrices with matrix sizes $N \times N$
and $M \times M$, and ``Tr'' and ``tr'' stand for the trace for them,
respectively.
This equality is just rephrased as
\begin{equation}
 e^{-\frac{1}{2g_s} \Tr A^2}
  \left\langle
   e^{\frac{1}{g_s} \Tr XA}
   \prod_{\alpha=1}^M
   \det( \lambda_\alpha - X )
  \right\rangle
  =
 (-1)^{MN}
 e^{\frac{1}{2g_s} \tr \Lambda^2}
  \left\langle
   e^{\frac{1}{g_s} \, i \, \tr Y \Lambda}
   \prod_{j=1}^N
   \det( a_j + i Y )
  \right\rangle
  \, .
  \label{Gaussian_duality02} 
\end{equation}
From the generic formula (\ref{dual_part_func02}), we can understand this
duality as a consequence of the self-duality of Hermite polynomials with
respect to Fourier transformation.
Actually Hermite polynomial, which is an orthogonal polynomial with the
Gaussian weight function $w(x) = e^{-\frac{1}{2}x^2}$, has an integral
representation
\begin{equation}
 H_k (x)
  = 
  \int_{-\infty}^\infty \frac{dt}{\sqrt{2\pi}} \,
  (it)^k \, e^{-\frac{1}{2}(t+ix)^2}
  \, .
\end{equation}
This expression is essentially the same as (\ref{aux_func02}) in the case of
the harmonic potential.
This is a specific property for the Gaussian model.

\subsection{Fermionic formula}\label{sec:fermionic}

For the Gaussian matrix model, there is another interesting derivation of
the duality (\ref{Gaussian_duality}) using fermionic
variables, instead of the method used in Sec.~\ref{sec:formula}.
Following the approach applied in~\cite{Brezin:2000CMP,Brezin:2007aa}
basically, we discuss it from the view point of the topological strings.
We will actually show that the effective fermionic action for this
partition function gives a quite natural perspective on the topological brane.

Introducing fermionic variables in bifundamental representations
$(N,\bar{M})$ and $(\bar{N},M)$ of $\U(N)\times\U(M)$, the
characteristic polynomial is involved in an exponential form
\begin{equation}
 \prod_{\alpha=1}^M \det (\lambda_\alpha - X)
  = \int \! d [\bar\psi, \psi] \,
  e^{\sum_{\alpha=1}^M \bar\psi^\alpha_i (\lambda_\alpha \id - X)_{ij}
  \psi^\alpha_j}
  \, .
\end{equation}
Using this formula, the LHS of (\ref{Gaussian_duality02}) becomes
\begin{equation}
 \frac{1}{\calZ_N} \int \! d [X, \bar\psi, \psi] \,
  e^{-\frac{1}{2g_s} \Tr (X-A)^2 + \bar\psi^\alpha_i (\lambda_\alpha \id
  - M)_{ij} \psi_j^\alpha} .
\end{equation}
Then the effective action yields
\begin{eqnarray}
 S_{\rm eff}(X, \bar\psi, \psi)
 & = &
  -\frac{1}{2g_s} \Tr (X-A)^2 
  + \bar\psi^\alpha_i (\lambda_\alpha \id - M)_{ij} \psi_j^\alpha
  \nonumber \\
% & = & -\frac{1}{2g_s} \Tr (M-A)^2 
%  + \Tr M \psi^\alpha \bar\psi^\alpha 
%  - \tr \Lambda \, \psi_j \bar \psi_j
%  \nonumber \\
 & = & -\frac{1}{2g_s} \Tr (X - A - g_s \psi^\alpha \bar\psi^\alpha)^2 
  + \Tr A \, \psi^\alpha \bar\psi^\alpha
  + \frac{g_s}{2} \psi^\alpha_i \bar\psi^\alpha_j \psi^\beta_j \bar\psi^\beta_i
  - \tr \Lambda \, \psi_j \bar \psi_j
  \, ,
  \nonumber \\
\end{eqnarray}
where 
%$\Lambda$ is a $k \times k$ diagonal matrix $\Lambda =
%\mathrm{diag}(\lambda_1, \cdots, \lambda_k)$, and $\tr$ stands for the
%trace for $k\times k$ matrix with indices $\alpha = 1, \cdots, k$.
$(\psi^\alpha \bar\psi^\alpha)_{ij} \equiv \psi^\alpha_i \bar\psi^\alpha_j$
and
$(\psi_j \bar\psi_j)^{\alpha\beta} \equiv \psi^\alpha_j \bar\psi^\beta_j$
are $N\times N$ and $M\times M$ matrices, respectively.
Integrating out the matrix $X$, we obtain the intermediate
form of the formula, which can be represented only in terms of the
fermionic variables
\begin{equation}
 e^{-\frac{1}{2g_s} \Tr A^2}
 \left\langle
  e^{\frac{1}{g_s} \Tr XA} 
  \prod_{\alpha=1}^M
  \det ( \lambda_\alpha - X )
 \right\rangle
 =
 \int \! d[\bar\psi,\psi] \,
  e^{
  \frac{g_s}{2} \psi^\alpha_i \bar\psi^\alpha_j \psi^\beta_j \bar\psi^\beta_i
  +\Tr A \, \psi^\alpha \bar\psi^\alpha
  - \tr \Lambda \, \psi_j \bar \psi_j}
  \, .
  \label{fermion_rep}
\end{equation}
Since the four-point interaction is also represented in terms of the $M
\times M$ matrix as $ \Tr \left(\psi^\alpha\bar\psi^\alpha\right)^2 = -
\tr \left(\psi_i \bar\psi_i\right)^2$,
%\begin{equation}
% \Tr \left(\psi^\alpha\bar\psi^\alpha\right)^2
%  = \psi^\alpha_i \bar\psi^\alpha_j \psi^\beta_j \bar\psi^\beta_i
%  = - \psi^\alpha_i \bar\psi^\beta_i \psi^\beta_j \bar\psi^\alpha_j
%  = - \tr \left(\psi_i \bar\psi_i\right)^2
%  \, ,
%\end{equation}
this term can be removed by inserting an $M\times M$ auxiliary matrix $Y$
\begin{eqnarray}
 & &
  \frac{1}{\calZ_M}
  \int \! d[Y, \bar\psi,\psi] \,
  e^{ - \frac{1}{2g_s} \tr (Y - i \Lambda - i g_s \psi_j \bar\psi_j)^2
  - \tr \Lambda \, \psi_j \bar \psi_j
  - \frac{g_s}{2} \tr \left(\psi_i \bar\psi_i\right)^2
  + \Tr A \, \psi^\alpha \bar\psi^\alpha }
  \nonumber \\
% & = &
%  \frac{1}{\calZ_M}
%  \int \! d[Y, \bar\psi,\psi] \,
%  e^{ - \frac{1}{2g_s} \tr (Y - i \Lambda )^2
%  + \Tr A \, \psi^\alpha \bar\psi^\alpha
%  + \tr i \, Y \, \psi_j \bar\psi_j }
%  \nonumber \\
 & = &
  \frac{(-1)^{MN}}{\calZ_M}
  \int \! dY \,
  \prod_{j=1}^N \det (a_j + i Y) \,
  e^{ - \frac{1}{2g_s} \tr (Y - i \Lambda )^2 }
%  \nonumber \\
% & = & (-1)^{MN} 
%  \left\langle
%   e^{\frac{1}{g_s} \, i \, \tr Y \Lambda}
%   \prod_{j=1}^N \det (a_j + i Y) \,
%  \right\rangle
  \, .
\end{eqnarray}
This is just the RHS of the duality formula
(\ref{Gaussian_duality02}).

\begin{figure}[t]
% \begin{center}
  \hspace{6em}
  \includegraphics[width=30em]{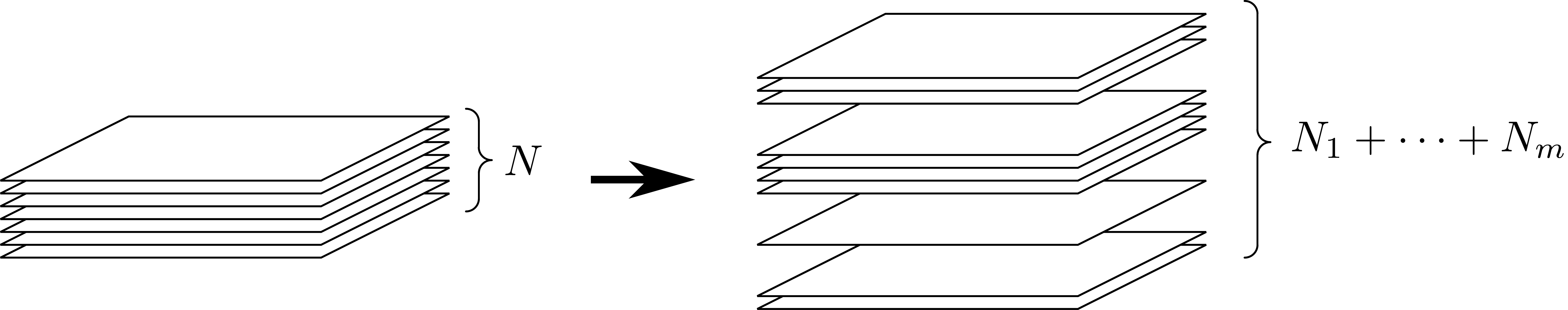}
% \end{center}
 \caption{When we turn on the external source $A$, degenerated
 positions of $N$~branes are lifted.
 Correspondingly, $\U(N)$ symmetry is broken into
 $\U(N_1)\times\cdots\times\U(N_m)$ by applying $m$ distinct values
 to $A$.}
 \label{brane_ext}
\end{figure}

Let us comment on the meaning of this formula in terms of the
topological strings.
When we apply $m$ distinct values to $A$ as
\begin{equation}
 A = \mathrm{diag}
  (
  \underbrace{a^{(1)}, \cdots, a^{(1)}}_{N_1}, 
  \underbrace{a^{(2)}, \cdots, a^{(2)}}_{N_2},
  \cdots\cdots,
  \underbrace{a^{(m)}, \cdots, a^{(m)}}_{N_m}
  )
  \, ,
\end{equation}
stacked $N$ branes are decoupled into $N_1+\cdots+N_m$ as shown in
Fig.~\ref{brane_ext}.
This means that the $\U(N)$ symmetry of the original matrix model is
broken into its subsector
\begin{equation}
 \U(N) \ \longrightarrow \
  \U(N_1) \times \cdots \times \U(N_m) 
  \, .
\end{equation}
We find a similar symmetry breaking in the dual representation.
In particular, when we put $\Lambda$ as
\begin{equation}
 \Lambda = \mathrm{diag}
  (
  \underbrace{\lambda^{(1)}, \cdots, \lambda^{(1)}}_{M_1}, 
  \underbrace{\lambda^{(2)}, \cdots, \lambda^{(2)}}_{M_2},
  \cdots\cdots,
  \underbrace{\lambda^{(l)}, \cdots, \lambda^{(l)}}_{M_l}
  )
  \, ,
\end{equation}
the $\U(M)$ symmetry is broken as
\begin{equation}
 \U(M) \ \longrightarrow \
  \U(M_1) \times \cdots \times \U(M_l) 
  \, .
\end{equation}

We can also discuss the symmetry breaking of the fermions by seeing the
fermionic effective action in (\ref{fermion_rep}),
\begin{equation}
 S_{\rm eff}(\psi,\bar\psi) = 
  \frac{g_s}{2} \psi^\alpha_i \bar\psi^\alpha_j \psi_j^\beta \bar\psi^\beta_i
  + \Tr A \, \psi^\alpha \bar \psi^\alpha
  - \tr \Lambda \, \psi_j \bar \psi_j 
  \, .
\end{equation}
Although the four-point interaction term is invariant under the full
symmetry of $\U(N)\times\U(M)$, this symmetry is partially broken, because the
source term plays a role of the non-singlet mass term.
The remaining symmetry is $\U(N_i)\times\U(M_j)$ for $i=1, \cdots, m$
and $j=1, \cdots, l$, as a subset of $\U(N)\times\U(M)$.
%It just corresponds to each intersection point in the schematic picture
%of the brane intersection (Fig.~\ref{brane_intersect02}).
This fermionic excitation should be seen as a remnant
of the chiral fermion associated with the anomaly on the intersecting
branes~\cite{Green:1996dd,Berkooz:1996km,Blumenhagen:2005mu,Itzhaki:2005tu,Dijkgraaf:2007sw}.

%\begin{figure}[t]
%% \begin{center}
% \hspace{3em}
%  \includegraphics[width=35em]{brane_intersect02}
%% \end{center}
% \caption{{\em Waffle structure of the intersecting branes (W-brane)}. 
% A massless fermionic excitation appears on each intersecting point of this
% configuration, which is in the bifundamental representation of
% $\U(N_i)\times\U(M_j)$ for $i=1, \cdots, m$ and $j=1, \cdots, l$.
% The $p$ and $x$ directions are converted into each other through the Fourier
% transformation.
% }
% \label{brane_intersect02}
%\end{figure}

\subsection{Bosonic formula}\label{sec:bosonic}

We can extend the duality formula (\ref{Gaussian_duality02}) for the
inverse characteristic polynomial~\cite{Desrosiers:2008tp}
\begin{equation}
 e^{-\frac{1}{2g_s} \Tr A^2}
 \left\langle
  e^{\frac{1}{g_s} \Tr XA}
  \prod_{\alpha=1}^M \det (\lambda_\alpha - X)^{-1}
 \right\rangle
 \ = \ 
 e^{-\frac{1}{2g_s} \tr \Lambda^2}
 \left\langle
  e^{\frac{1}{g_s} \tr Y\Lambda}  
  \prod_{j=1}^N \det (a_j - Y)^{-1}
 \right\rangle
 \, .
 \label{dual_rep_boson}
\end{equation}
In this case a bifundamental bosonic field plays a similar role to the
fermionic field, which is used to represent the characteristic
polynomial in the numerator.
Actually we can derive this duality formula in almost the same manner
as that discussed in Sec.~\ref{sec:fermionic}.

The average shown in the LHS of (\ref{dual_rep_boson}) is explicitly
written as
\begin{equation}
 e^{-\frac{1}{2g_s} \Tr A^2}
 \left\langle
  e^{\frac{1}{g_s} \Tr XA}
     \prod_{\alpha=1}^M \det (\lambda_\alpha - X)^{-1}
    \right\rangle
 = \frac{1}{\calZ_N} \int \! dX \,
 \prod_{\alpha=1}^M \det (\lambda_\alpha - X)^{-1}
 \, e^{-\frac{1}{2g_s} \Tr (X-A)^2} 
 \, .
\end{equation}
Since the inverse of a determinant is written as a Gaussian integral with
a bosonic variable in the bifundamental representations,
\begin{equation}
 \prod_{\alpha=1}^M \det (\lambda_\alpha - X)^{-1}
  = \int \! d [\phi, \phi^\dag] \,
  e^{- \sum_{\alpha=1}^M \phi^{*\alpha}_i (\lambda_\alpha \id - X)_{ij}
  \phi^\alpha_j},
\end{equation}
the intermediate form, corresponding to (\ref{fermion_rep}), is given by
\begin{equation}
% F(a_1, \cdots, a_N; \lambda_1, \cdots, \lambda_k) = 
 \int \! d[\phi,\phi^\dag] \,
  e^{
  \frac{g_s}{2} \phi^\alpha_i \phi^{*\alpha}_j \phi^\beta_j \phi^{*\beta}_i
  +\Tr A \, \phi^\alpha \phi^{*\alpha}
  - \tr \Lambda \, \phi_j \phi^*_j}.
  \label{boson_rep}
\end{equation}
Then, inserting an auxiliary $M \times M$ Hermitian matrix $Y$ in order
to eliminate the four-point interaction, it is written as
\begin{eqnarray}
 & &
  \frac{1}{\calZ_M}
  \int \! d[Y, \phi,\phi^\dag] \,
  e^{ - \frac{1}{2g_s} \tr (Y - \Lambda + g_s \phi_j \phi^*_j)^2
  - \tr \Lambda \, \phi_j \phi_j^*
  + \frac{g_s}{2} \tr \left(\phi_i \phi^*_i\right)^2
  + \Tr A \, \phi^\alpha \phi^{*\alpha} }
  \nonumber \\
% & = &
%  \frac{1}{\calZ_k}
%  \int \! \calD[B, \phi,\phi^\dag] \,
%  e^{ - \frac{1}{2g_s} \tr (B - \Lambda )^2
%  + \Tr A \, \phi^\alpha \phi^{*\alpha}
%  - \tr B \, \phi_j \phi^*_j }
%  \nonumber \\
 & = & 
  e^{-\frac{1}{2g_s} \tr \Lambda^2}
  \left\langle
   e^{\frac{1}{g_s} \tr Y \Lambda}
   \prod_{j=1}^N \det (a_j - Y)^{-1}
  \right\rangle
  \, .
\end{eqnarray}
This is the RHS of the duality formula (\ref{dual_rep_boson}).

The symmetry of the bifundamental bosons are partially
broken, as seen in the bosonic effective action (\ref{boson_rep}), as
well as the fermionic case.
The difference from the previous case is the role of branes.
From the topological string point of view, the inverse of the
characteristic polynomial is regarded as the partition function of the
non-compact anti-brane, which is created by the vertex operator with
the opposite charge $\overline{V}(x) =
e^{\frac{1}{2g_s}\phi(x)}$, instead of (\ref{vertex_op}).
In this case, the correlation function is written in terms of the Cauchy
transform of the corresponding orthogonal
polynomial~\cite{Fyodorov:2002jw}.

\section{Discussion}\label{sec:discussion}

In this article we have investigated the symplectic invariance of branes in the
topological B-model using
its matrix model description.
In particular, since two different descriptions of the non-compact brane
correspond to the characteristic polynomial and the external source in
the matrix model, we have considered the brane partition function given
by inserting both of them simultaneously.
We have derived the determinantal formula for this partition
function, and shown that two descriptions of the branes are dual to each
other in the sense of the Fourier transformation.
We have also shown that the brane partition function plays a role of the
$\tau$-function, and satisfies the Toda lattice equations in one and two
dimensions.
We have investigated the Gaussian matrix model as an example, and
discussed the effective action of the topological branes in terms of
the bifundamental fermion/boson.

Although we have focused on the $\U(N)$ symmetric matrix model
all through this article, we can apply essentially the same argument
to $\mathrm{O}(N)$ and $\Sp(2N)$ symmetric matrix models.
In such a case the Hermitian matrix is replaced with real symmetric and
self-dual quaternion matrices, respectively.
Actually, when the Gaussian potential is assigned, one can obtain a similar
duality formula~\cite{Brezin:2001CMP,Desrosiers:2008tp,Forrester:2013JPA},
which claims that the dual of the $\mathrm{O}(N)$ model is the
$\Sp(2N)$ model and vice versa.
This relation is extended to arbitrary $\beta$-ensemble, and generic
duality between $\beta$ and $1/\beta$ is found.
From the string theoretical point of view, this property is naturally
understood as insertion of orientifold plane.
Correspondingly the $\U(N)\times\U(M)$ bifundamental fermion/boson used
in Sec.~\ref{sec:Gaussian} is replaced with the $\mathrm{O}(N)
\times \Sp(2M)$ bifundamental variables.

Let us comment on some possible applications of the duality discussed in
this article.
The exact low energy dynamics of $\N=2$ gauge theory, which is described
by Seiberg--Witten theory, is solved by the world volume theory of
D4-branes suspended between NS5-branes, and also an M5-brane, appearing
in its M-theory lift~\cite{Witten:1997sc}.
In this case, since the geometry of this M5-brane indicates the Seiberg--Witten
curve of the corresponding $\N=2$ theory, positions of branes are
directly related to the gauge theory dynamics.
Thus it is expected that a nontrivial gauge theory duality is derived
from the duality between the two coordinates of branes.
Actually a similar duality is discussed along this direction~\cite{Bao:2011rc}.

From the matrix model perspective, it is interesting to consider the
ratio of the characteristic polynomials in the presence of the external
source~\cite{Fyodorov:2002jw}, and its interpretation in terms of
topological strings.
The characteristic polynomial in the numerator and the denominator plays
a role of the creation operator for the brane and anti-brane.
Thus the ratio should describe the pair creation and annihilation of
branes.
In particular it is expected that the scaling limit of the ratio
extracts some interesting features of the tachyon condensation in
topological strings.
This kind of problem is also interesting in the context of the matrix
model itself, because one can often find an universal property of the matrix
model in such a scaling limit.

\subsection*{Acknowledgements}

I would like to thank Bertrand Eynard for his insightful comment on the
symplectic invariance of the spectral curve.
I am also grateful to Masato Taki for valuable discussions on topological
string theory.
This work is supported in part by Grant-in-Aid for JSPS Fellows~(\#25-4302).
Finally it is my pleasure to submit the first version of this article on
my 30th birthday.
I would like to express my sincere gratitude to my parents on this
occasion, and dedicate this article to the memory of them.

\appendix
\section{A formula for determinants}\label{sec:app}

In order to obtain (\ref{R_relation01}) and (\ref{R_relation02}), it is
convenient to use the following relation
\begin{equation}
% \det B_M = M \, x \det A_M
 \frac{\det B_M(x)}{\det A_M(x)} = M \, x
  \, ,
  \label{shift_det}
\end{equation}
where we have introduced two $M \times M$ matrices
\begin{equation}
 A_M
  = 
 \left(
  \begin{array}{cccc}
   \left(x^{N}\right)^{(0)} & \cdots & \left(x^{N+M-2}\right)^{(0)} & 
    \left(x^{N+M-1}\right)^{(0)} \\
   \left(x^{N}\right)^{(1)} & \cdots & \left(x^{N+M-2}\right)^{(1)} & 
    \left(x^{N+M-1}\right)^{(1)} \\
   \vdots & & \vdots & \vdots \\
   \vdots & & \vdots & \vdots \\
%   \left(x^{N}\right)^{(M-2)} & \cdots & \left(x^{N+M-2}\right)^{(M-2)} & 
%    \left(x^{N+M}\right)^{(M-2)} \\
   \left(x^{N}\right)^{(M-1)} & \cdots & \left(x^{N+M-2}\right)^{(M-1)} & 
    \left(x^{N+M-1}\right)^{(M-1)} \\
  \end{array}
 \right)
 \, ,
\end{equation}
and
\begin{equation}
 B_M
  =
 \left(
  \begin{array}{cccc}
   \left(x^{N}\right)^{(0)} & \cdots & \left(x^{N+M-2}\right)^{(0)} & 
    \left(x^{N+M}\right)^{(0)} \\
   \left(x^{N}\right)^{(1)} & \cdots & \left(x^{N+M-2}\right)^{(1)} & 
    \left(x^{N+M}\right)^{(1)} \\
   \vdots & & \vdots & \vdots \\
   \vdots & & \vdots & \vdots \\
%   \left(x^{N}\right)^{(M-2)} & \cdots & \left(x^{N+M-2}\right)^{(M-2)} & 
%    \left(x^{N+M}\right)^{(M-2)} \\
   \left(x^{N}\right)^{(M-1)} & \cdots & \left(x^{N+M-2}\right)^{(M-1)} & 
    \left(x^{N+M}\right)^{(M-1)} \\
  \end{array}
 \right)
 \, .
\end{equation}
Here we denote $(x^j)^{(l)} = (d/dx)^l x^j$ and so on.
Using the Jacobi identity (\ref{Jacobi_id}) for $\det A_M$ with $i = k = M$
and $j = l = M - 1$, we have
\begin{equation}
 \det A_M \cdot \det A_{M-2}
  =
    \det A_{M-1} \cdot \partial_x \det B_{M-1}
  - \partial_x \det A_{M-1} \cdot \det B_{M-1}
  \, .
\end{equation}
It is convenient to rewrite this relation as
\begin{equation}
 \frac{\det A_M \cdot \det A_{M-2}}{\left(\det A_{M-1}\right)^2}
  =
  \frac{\partial}{\partial x}
  \left( \frac{\det B_{M-1}}{\det A_{M-1}} \right)
  \, .
  \label{Toda_poly}
\end{equation}
This is interpreted as a remnant of the Toda lattice
equation~\cite{Kharchev:1991cy}.

What we have to do next is to evaluate $\det A_M$.
To obtain this, we consider a ratio of determinants,
and then take the equal parameter limit, 
$x_\alpha \to x$ for all $\alpha = 1, \cdots, M$,
\begin{equation}
 \frac{1}{\Delta(x)}
  \det_{1 \le \alpha, \beta \le M} 
  \left( x_\beta \right)^{N+\alpha-1}
  =
% \frac{1}{\Delta(x)}
 \prod_{\alpha=1}^M x^N_\alpha 
%  \det_{1 \le \alpha, \beta \le M} 
%  \left( x_\beta \right)^{\alpha-1}
 \ \stackrel{x_\alpha \to x}{\longrightarrow} \
 x^{MN}
 \, .
\end{equation}
On the other hand, it is written in another form
\begin{equation}
 \frac{1}{\Delta(x)}
 \det_{1 \le \alpha, \beta \le M} 
 \left( x_\beta \right)^{N+\alpha-1}
 \ \stackrel{x_\alpha \to x}{\longrightarrow} \
  \left(
   \prod_{\alpha=1}^M (\alpha-1)!
  \right)^{-1}
 \det_{1 \le \alpha, \beta \le M} 
 \left( x^{N+\alpha-1} \right)^{(\beta-1)}
 \, .
\end{equation}
Comparing these two expressions, we obtain the following result
\begin{equation}
 \det A_M 
  =
  \left(
   \prod_{\alpha=1}^M (\alpha-1)!
  \right)
  x^{MN}
  \, .
\end{equation}
Substituting this expression into the relation (\ref{Toda_poly}), we
arrive at the relation (\ref{shift_det}).

%%%%%%%%%  Rerefence  %%%%%%%%%

\bibliographystyle{ytphys}

%\bibliography{/home/kimura/Configure/conf}
\bibliography{/Users/k_tar/Dropbox/etc/conf}

\end{document}